\documentclass[aps,nofootinbib,twocolumn,floatfix]{revtex4}

\usepackage{graphicx}
\usepackage{amsmath}
\usepackage{bm} 
\usepackage{color}

\newcommand{\beq}{\begin{eqnarray}}
\newcommand{\eeq}{\end{eqnarray}}
\newcommand{\bqa}{\begin{eqnarray}}
\newcommand{\eqa}{\end{eqnarray}}

\def\mqo2{{\!\!\!}}
\renewcommand{\vec}{\mathbf}

\begin{document}


\preprint{HISKP-TH-09/xx}
\title{Resonant Atom-Dimer Relaxation in Ultracold Atoms}
\author{Kerstin Helfrich}
\author{H.-W. Hammer}
\affiliation{~\\Helmholtz-Institut f\"ur Strahlen- und Kernphysik (Theorie)\\
and Bethe Center for Theoretical Physics,
 Universit\"at Bonn, 53115 Bonn, Germany\\}

\date{\today}


\begin{abstract}
Three-body systems with large scattering length display 
universal phenomena associated with a discrete scaling symmetry.
These phenomena include resonant enhancement of three-body loss rates
when an Efimov three-body resonance is at the scattering threshold. 
In particular,
there can be resonant peaks in the atom-dimer relaxation rate 
for large positive scattering length.
We improve upon earlier studies and calculate the atom-dimer relaxation rate
as a function of temperature using a Bose-Einstein distribution for the
thermal average. As input, we use calculations of the 
atom-dimer scattering phase shifts from effective field theory.
\end{abstract}

\maketitle


%
{\it Introduction}:
The Efimov effect is a phenomenon that occurs in a nonrelativistic
3-body system with resonant short-range interactions.
Efimov discovered in 1970 that there are 
infinitely many 3-body bound states with an accumulation point at the 
scattering threshold when the S-wave scattering length $a$ is tuned 
to the unitary limit $1/a =0$ \cite{Efimov70}: 
\begin{eqnarray}
E^{(n)}_T = (e^{-2\pi/s_0})^{n-n_*} \hbar^2 \kappa^2_* /m,
\label{kappa-star}
\end{eqnarray}
where $m$ is the mass of the particles, $s_0 \approx 1.00624$, and
$\kappa_*$ is the binding wavenumber of the branch of Efimov states 
labeled by $n_*$.  The geometric spectrum in (\ref{kappa-star})
is the signature of a 
discrete scaling symmetry with scaling factor $e^{\pi/s_0}\approx 22.7$.
For a finite scattering length that is large compared to the range of the 
interaction, the universal properties persist but
there will only be a finite number of Efimov states. The corrections 
to the unitary limit are calculable in perturbation
theory \cite{Efimov:1993zz,Hammer:2006zs,Platter:2008cx}.
The Efimov effect is just one example of universal
phenomena in the 3-body system with large scattering length 
\cite{Efimov71,Efimov79}. For reviews of this \lq\lq Efimov physics'', 
see Refs.~\cite{Braaten:2004rn,Braaten:2006vd}.
Here we focus on identical bosons in a single spin state and positive 
scattering length.

Since we are interested in applications to ultracold atoms,
we refer to the bosons as atoms, 
their 2-body bound states as dimers, 
and their 3-body bound states as trimers.  
If the scattering length $a$ is large and positive, there is a  
shallow dimer with binding energy $E_D=\hbar^2/(ma^2)$.
Moreover, there is an infinite sequence of
positive values of $a$ for which there is an Efimov trimer at the 
atom-dimer scattering threshold \cite{Efimov71}:
$a=(e^{\pi/s_0})^n a_*$, where $a_* \approx 0.0708 \kappa_*^{-1}$
\cite{BH02}. 
Another example of Efimov physics is an infinite sequence
of positive values of $a$ for which the 3-body recombination rate 
into the shallow dimer vanishes \cite{NM-99,EGB-99,BBH-00,BH01}. 
The universal aspects of Efimov physics are 
determined by two parameters: the scattering length $a$
and the Efimov parameter $\kappa_*$.

The alkali atoms used in most ultracold atom experiments 
have many deeply-bound diatomic molecules (deep dimers).
Efimov physics is modified by the existence of the deep dimers
because they provide inelastic channels for scattering processes and 
the decay of trimers. If there is an Efimov trimer 
near the scattering threshold, there will be
a resonant enhancement of inelastic scattering processes.
When $a \approx (e^{\pi/s_0})^n a_*$, the resonant inelastic process 
is dimer relaxation, in which the collision of an atom 
and a shallow dimer produces an atom and a deep
dimer \cite{Braaten:2003yc}.
If there are deep dimers, the universal phenomena associated 
with Efimov physics are determined by three parameters:
$a$, $\kappa_*$, and a parameter $\eta_*$ that determines 
the widths of Efimov trimers \cite{Braaten:2003yc}.

Experimental evidence for an Efimov trimer in ultracold Cs
atoms was recently provided by their signature in three-body
recombination rates \cite{Kraemer-06}. This signature could be
unravelled by varying the scattering length $a$ over several orders of
magnitude using a Feshbach resonance.  More recently, 
possible evidence for an Efimov resonance was also obtained in 
atom-dimer scattering \cite{Knoop08}, 
in three-body recombination in a balanced mixture of 
atoms in three different hyperfine states of 
$^6$Li~\cite{Ottenstein08,Huckans08}, and in a 
heteronuclear system of $^{41}$K and $^{87}$Rb~\cite{Barontini09}.

In this paper, we focus on the Efimov resonance in atom-dimer scattering
found in \cite{Knoop08}.
This process was previously considered in \cite{Braaten:2003yc,Braaten:2006nn}.
We go beyond these earlier studies in various respects: we use the
full effective field theory results for the atom-dimer phase shifts instead 
of the effective range expansion and perform a thermal average using the
Bose-Einstein instead of the Boltzmann distribution. Moreover, we correct 
an error in the calculation of \cite{Braaten:2006nn}.

{\it Atom-dimer scattering}: We consider the scattering of an atom with mass
$m_A=m$ and dimer with mass $m_D=2m$.
The wavenumbers $\vec{p}_A$ and $\vec{p}_D$ of the incoming atom and
dimer, respectively, can be decomposed into the 
total wavenumber $\vec{p}_{\rm tot}=\vec{p}_A+\vec{p}_D$ and the relative 
wavenumber $\vec{k}=\frac{2}{3} \vec{p}_A-\frac{1}{3} \vec{p}_D$. 
Because of Galilei invariance, the scattering observables depend
on the relative wavenumbers and the collision energy $E$ in the
center-of-mass system only: $E=3 \hbar^2 k^2/(4 m)$ where $k=|\vec{k}|$. 
The differential cross section for elastic 
atom-dimer scattering is
\begin{eqnarray}
\frac{d\sigma_{AD}^{\rm (elastic)}}{d\Omega} = \left|f_{AD}(k,\theta)\right|^2\,,
\label{diffcross}
\end{eqnarray}
where $f_{AD}(k,\theta)$ is the scattering amplitude. The elastic cross
section $\sigma_{AD}^{\rm (elastic)}$ is obtained by integrating 
Eq.~(\ref{diffcross}) over the full solid angle.
The total cross section (including elastic and inelastic contributions)
can be calculated using the optical theorem:
\begin{eqnarray}
\sigma_{AD}^{\rm (total)} = \frac{4 \pi}{k}{\rm Im} f_{AD}(k,\theta=0)\,,
\label{totcross}
\end{eqnarray}
such that the inelastic cross section is given by the difference of the
total and elastic cross section.
At low energies, higher partial waves with $L>0$ are suppressed and
the scattering amplitude is dominated by S-waves ($L=0$): 
\begin{eqnarray}
f_{AD}(k) = \left[ k \cot \delta_0^{AD}(k)-ik\right]^{-1}\,.
\label{amplit}
\end{eqnarray}
For the S-wave atom-dimer phase shift $k \cot \delta_0^{AD}(k)$,
we will use the results from a calculation using the effective field 
theory of Ref.~\cite{Bedaque:1998kg}.  
A convenient parametrization of these results was given 
in~\cite{BH02,Braaten:2004rn}:
\begin{eqnarray}
\label{phaseshift}
ka \cot\delta_0^{AD}(k)&=&c_1(ka)\\ \nonumber
&+&c_2(ka) \cot\bigl[s_0\ln\bigl(0.19\, a/a_*\bigr)+\phi(ka)\bigr]\,,
\end{eqnarray}
where
\begin{eqnarray}
c_1(ka)& = &-0.22+0.39k^2a^2-0.17k^4a^4\,, \nonumber\\
c_2(ka)& = &0.32+0.82k^2a^2-0.14k^4a^4\,, \nonumber\\
\phi(ka)& = &2.64-0.83k^2a^2+0.23k^4a^4\,.
\label{parametrization}
\end{eqnarray}
This parametrization is valid up to the dimer breakup 
wavenumber of $k_{D}=2/(\sqrt{3} a)$.
A parametrization for higher wavenumbers beyond the dimer breakup
exists \cite{Braaten:2008kx} but will not be required for our 
purposes as we will demonstrate below. 

To leading order in the large scattering length, atom-dimer relaxation
can only proceed via S-waves. For the relaxation into deep dimers to
take place, the atom and the dimer have to approach each other to 
very short distances. However, because of the angular momentum barrier
this can only happen in the relative S-wave. The parametrization of
the S-wave phase shift in Eqs.~(\ref{phaseshift}, \ref{parametrization})
is therefore sufficient to calculate atom dimer-relaxation.

{\it Atom-dimer relaxation}: To incorporate the effects of deep dimers, 
we make the simple replacement in the amplitude~\cite{Braaten:2003yc}
\begin{eqnarray}
\ln a_* \rightarrow \ln a_*-i\eta_*/s_0\,,
\label{eta-star}
\end{eqnarray}
where $\eta_*$ determines the probability for an atom and a dimer
to scatter into an energetic atom and deep dimer pair at short
distances. This inelastic process generates 
the width of the Efimov resonances. 
The phase shift becomes imaginary even 
below the dimer breakup threshold and the released binding energy 
is converted 
to the kinetic energy of the recoiling atom and dimer. They thus are lost 
to the system. 
The event rate $\beta$ for this dimer relaxation process in an ultracold 
gas of atoms and dimers can be written as
\begin{eqnarray}
\frac{d}{dt}n_A=\frac{d}{dt}n_D=-\beta n_A n_D\,,
\label{betadef}
\end{eqnarray}
where $n_A$ and $n_D$ denote the number density of the atoms and dimers,
respectively. 

For an ensemble of atoms and dimers at
nonzero temperature that are held in a trap, temperature and 
trap geometry have to be included in the calculation of the observed dimer 
losses. The dimer loss rate can be expressed as
\begin{eqnarray}
\frac{d}{dt}N_D=-\int d^3r\,
\prod_{i=A,D}\left[\int\frac{d^3p_i}{(2\pi)^3}n_i(p_i,r)\right]g(k),
\label{ndotdef}
\end{eqnarray}
where $N_D$ is the number of dimers and
we use the generalized Bose-Einstein distribution function 
\begin{eqnarray}
n_i(p_i,r)=\biggl[\exp\Bigl\{\Bigl(\frac{\hbar^2 p_i^2}{2m_i}+
\frac{m_i\overline{\omega}^2r^2}{2}-\mu_i\Bigr)/k_B T\Bigr\}-1\biggr]^{-1},
\label{BEdef}
\end{eqnarray}
with $i=A,D$ denoting an atom or dimer, respectively.
The properties of the trap enter via the  average trap frequency
$\overline{\omega}$ while the function $g(k)$ to be averaged is given by
\begin{eqnarray}
g(k)&=&\frac{3 \hbar k}{2m}\,\sigma^{({\rm inel.})}_{AD}(k)\nonumber\\
&=&\frac{3 \hbar k}{2m}\left( \sigma^{({\rm total})}_{AD}(k)
-\sigma^{({\rm elastic})}_{AD}(k)\right)\,.
\label{sigmain}
\end{eqnarray}
In the limit $k\to 0$, $g(k)$ reduces to the relaxation rate constant
at zero temperature.
Note that the function $g(k)$ introduces an implicit dependence on the 
angle between $\vec{p}_{A}$ and $\vec{p}_D$. 
The chemical potentials $\mu_i$ are fixed via the equation 
\begin{eqnarray}
\int d^3 r\int \frac{d^3p_i}{(2 \pi)^3}\, n_i(p_i, r)=N_i\,,
\label{partnumber}
\end{eqnarray}
with $N_i$ being the particle number and $i=A,D$.

All angular integrations except for one can be carried out immediately
and the expression (\ref{ndotdef}) can be rewritten as:
\begin{eqnarray}
\frac{d}{dt}{N}_D &=&- ·\frac{1}{2\pi^3}\int_0^\infty r^2 dr\int_0^\infty 
p_{tot}^2 dp_{tot}  \int_0^\infty k^2 dk \nonumber\\ 
&& \times \int_{-1}^1 dx  \, n_A(p_A,r) n_D(p_D,r) g(k)\,,
\label{ndotT}
\end{eqnarray}
where $x$ is the cosine of the angle between $\vec{p}_{tot}$ and $\vec{k}$.

In evaluating Eq.~(\ref{ndotT}), we will cut off the integral over $k$
at the breakup wavenumber $k_{D}=2/(\sqrt{3} a)$ since the parametrization
in Eqs.~(\ref{phaseshift}, \ref{parametrization}) is only valid up to
$k_D$. We have estimated the error from this simplification by using
the unitary bound $i/k$ for the S-wave scattering amplitude $f_{AD}(k)$.
For the data of  Ref.~\cite{Knoop08},
the error involved is largest for the largest scattering length considered 
but even there only adds up to $0.2\%$ for the largest temperature
$T=170$ nK.
As a consequence, we can simply neglect the contribution from
$k>k_D$ in the analysis of the data.
\begin{figure}[t]
        \vspace*{0.4cm}
	\centerline{\includegraphics*[width=8cm]{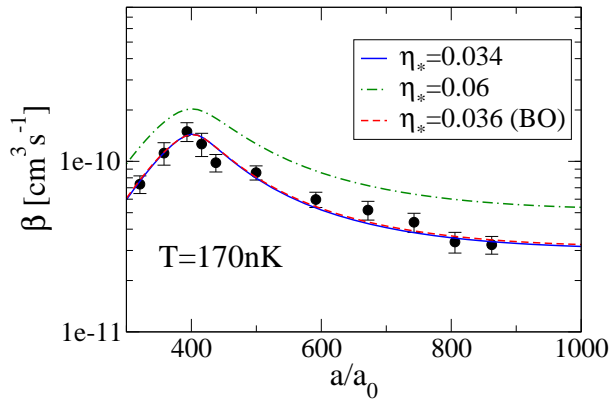}}
	\caption{The dimer relaxation coefficient $\beta$ as a function of 
$a/a_0$ for $T=170$~nK, $a_*=397 a_0$, and different values of 
$\eta_*$. The data points are from~\cite{Knoop08}. 
BO indicates a Boltzmann average.}
	\label{fig:01}
\end{figure}

{\it Results and Discussion}: 
We now apply our formalism to the experimental data
for the atom-dimer relaxation rate of ultracold Cs atoms 
as a function of the scattering length obtained by Knoop et al. 
\cite{Knoop08}.  Our free parameters are $a_*$ which determines the 
position of the resonance and $\eta_*$ which determines its width.
These parameters cannot be calculated in our approach and must be
taken from experiment. We will determine $a_*$ and  $\eta_*$ from the data
of Knoop et al.~and compare our results with what is 
known from other experiments.
In Ref.~\cite{Knoop08}, the dimer relaxation coefficient was extracted from
the dimer loss data using a loss model resulting in the rate equation
\begin{equation}
\frac{d}{dt}N_D=-\frac{8}{\sqrt{27}}\,\beta\, \bar{n}_A N_D +{\rm\ dimer\ 
loss\ term}\,,
\end{equation}
with $\bar{n}_A=[m\overline{\omega}^2/(4\pi k_B T)]^{3/2}N_A$ the mean 
atomic density. In order to compare our calculation with the experiment
of Knoop et al., we extract a value for $\beta$ from our result for $dN_D/dt$ (cf.~Eq.~(\ref{ndotT})) using
\begin{equation}
\beta \equiv -\frac{\sqrt{27}}{8 \bar{n}_A N_D} \frac{d}{dt}N_D\,.
\end{equation}

We start with the data at $T=170$~nK and fix the chemical potentials as 
described above. For atom number $N_A=10^5$, dimer number $N_D=4\times 10^3$, 
and an average trap frequency of 
$\overline{\omega}=45$~Hz~\cite{Knoop08pc}, we obtain the chemical potentials 
$\mu_A=-2.74\times 10^{-7}\ k_B$K and $\mu_D=-8.17\times 10^{-7}\ k_B$K.
In Fig.~\ref{fig:01} the data for the recombination constant $\beta$ is shown 
together with our best fit as the solid (blue) line. 
We only take into account data points for $a>300 a_0$. 
This fit yields $\chi^2/dof=1.2$.
We obtain for the peak position $a_*=397 a_0$ and for the 
resonance width parameter $\eta_*=0.034$. 
Also shown as a dashed-dotted (green) line is the resulting 
curve for the same resonance position but with $\eta_*=0.06$. 
This value of $\eta_*$ was obtained from fitting the three-body recombination
resonance in Cs for negative scattering length in~\cite{Kraemer-06}. It is
also compatible with the three-body recombination data for positive scattering
length presented in the same paper.\footnote{Note that in the Cs experiment
of \cite{Kraemer-06} the regions $a>0$ and $a<0$ are not required to have
the same parameters since they are separated by a zero in the scattering
length rather than a pole. See also the discussion in \cite{Knoop08}.}
However, the data for $a>0$  
are not very sensitive to the precise value of $\eta_*$
and values of $\eta_*$ as small as 0.01 
would also be compatible.
The width parameter  $\eta_*$ should only be weakly dependent on the magnetic
field in a universal region \cite{Braaten:2004rn}. 
In first approximation it can be assumed to
remain constant. A more serious puzzle is that the resonance position
$a_*=397 a_0$ extracted from the dimer relaxation data 
is not compatible with the value $a_*\approx 1200 a_0$
found in \cite{Kraemer-06}. The three-body recombination data 
cannot be satisfactorily described using $a_*=397 a_0$.
This disagreement requires
further study. However, one has to keep in mind that the atom-dimer
resonance is at the border of the universal region since the van der Waals 
length scale is $l_{vdW}\approx 200 a_0$ for Cs atoms. 

For temperatures much larger than the Bose-Einstein-condensation 
temperature, the thermal average can be replaced by a Boltzmann 
average. 
The critical temperatures are estimated by setting the chemical potential 
to zero and solving Eq.~(\ref{partnumber}) for the critical temperature. 
This yields $T_{c,A}\approx 94$~nK for the atoms and  $T_{c,D}\approx 32$~nK 
for the dimers. Therefore, it 
seems justified to use Boltzmann distributions instead of the 
Bose-Einstein distribution functions $n_i$. The resulting calculation 
for $\beta$ is analogous to the method of Ref.~\cite{Braaten:2006nn} 
but uses the parametrization of the scattering phase (\ref{phaseshift})
instead of an effective range expansion. The resulting curve is also 
shown in Fig.~\ref{fig:01} as the dashed (red) line. We remark that in the 
numerical evaluation of Eq.~(8) in
Ref.~\cite{Braaten:2006nn} a factor of $k^2$ in the thermal average
was omitted. This lead to a wrong normalization of the curves 
in Figs.~1 and 2 of this reference. The dashed line obviously 
also describes the data quite well but yields $\eta_*=0.036$. This 
shows how temperature dependence and averaging methods can partly 
be compensated by adjusting the width parameter $\eta_*$. For an 
accurate determination of $\eta_*$, reliable temperature and particle number 
measurements are thus crucial.

Using the effective range expansion for the atom-dimer scattering
amplitude as in Ref.~\cite{Braaten:2006nn} instead of the phase shift
parametrization of Eqs.~(\ref{phaseshift}, \ref{parametrization}) does 
not alter the overall shape or normalization of the dimer relaxation 
coefficient $\beta$. However, the extracted value of $a_*$ is shifted
by about 3\% to a higher value whereas $\eta_*$ remains unchanged.
The scattering length approximation with the atom-dimer scattering 
length given by $a_{AD}/a=1.46+2.15 \cot[s_0 \ln(a/a_*)+i\eta_*]$
does not lead to an equally good fit result. 
The obtained values for $\beta$ are 
smaller especially for higher values of $a$. Besides a change in peak position,
width and height, we can only obtain a $\chi^2/dof$ of $3-4$.

\begin{figure}[t]
   \vspace*{0.4cm}
	\centerline{\includegraphics*[width=8cm]{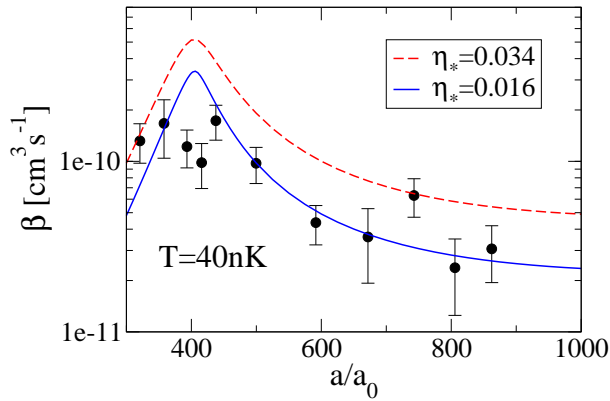}}
	\caption{The dimer relaxation coefficient $\beta$ as a function of 
$a/a_0$ for $T=40$~nK, $a_*=397 a_0$, and different values of 
$\eta_*$. The data points are from~\cite{Knoop08}. 
}
	\label{fig:02}
\end{figure}
We now turn to the data for $T=40$~nK~\cite{Knoop08} and compare 
them to our theoretical results for different values of $\eta_*$.
Here, the atom and dimer numbers are
$N_A=3\times 10^4$ and $N_D=4\times10^3$, and the trap frequency is 
$\overline{\omega}=25.2$~Hz \cite{Knoop08}
leading to the chemical potentials 
$\mu_A=-1.21\times 10^{-8}\ k_B$K and
$\mu_D=-8.87\times 10^{-8}\ k_B$K.
The critical temperatures are estimated as $T_{c,A}\approx 35$~nK
for the atoms and $T_{c,D}\approx 18$~nK for the dimers such 
that the temperature is only slightly 
larger than the critical temperature for the 
atoms. In Fig.~\ref{fig:02} we show the data for the relaxation coefficient 
$\beta$ at $T=40$~nK together with our fit results.
The dashed (red) line gives our prediction for 
the relaxation coefficient $\beta$ using the parameters obtained by fitting 
the $170$~nK data. The prediction is about a factor 2 too large 
compared to the data.
The dip in the data at the peak position
represented by the third and fourth  data point
cannot be reproduced within our theory.
If it is not simply a statistical fluctuation, 
it must be due to physics not captured in our theory such as 
non-universal effects or four-body physics
\cite{Platter:2004qn,Hammer:2006ct,vStech08}.
If we keep the resonance position at $a_*=397 a_0$ but 
fit the parameter $\eta_*$ to the 40~nK data excluding the third and fourth 
data point, we obtain the solid (blue) line. 
Still excluding the third and fourth 
data point this gives $\eta_*=0.016$ with $\chi^2/dof=1.5$ and describes 
the data satisfactorily. 

In summary, we have calculated the atom-dimer relaxation rate 
for large positive scattering length in a universal zero-range
approach. We have improved on previous
studies  \cite{Braaten:2003yc,Braaten:2006nn} by using a Bose-Einstein 
distribution for the thermal average and calculations of the 
atom-dimer scattering phase shifts from effective field theory.
Our results were then applied to the atom-dimer relaxation data
for Cs obtained by Knoop et al.~\cite{Knoop08}. Fitting the 
resonance position and width parameters $a_*$ and $\eta_*$, we could get
a good description of the $170$ nK data. Using these parameters, we
overpredict the relaxation data at $40$ nK by a factor of two.
Moreover, our theory is not able to reproduce the dip in the 40 nK
data directly at the resonance position and the corresponding 
physics appears to be missing in our theory.
We demonstrated that this discrepancy is neither due to the thermal averaging
procedure nor due to the phase shift parametrization used. 
The resonance position at the border of the universal region and the
mismatch in the extracted resonance position from atom-dimer relaxation
and the three-body recombination data \cite{Kraemer-06}
suggests that non-universal physics could be responsible.
However, it is also conceivable that four-body losses become
important at the lower temperature. This question deserves further study.

We thank Eric Braaten, Francesca Ferlaino, and Steven Knoop 
for discussions and Steven Knoop for providing his data. 
HWH was supported by the BMBF under contract No.~06BN411. K.H. was 
supported by the \lq\lq Studien\-stiftung des 
Deutschen Volkes'' and by the
Bonn-Cologne Graduate School of Physics and Astronomy.


\end{document}